# Dislocation breakaway from nanoparticle array linear complexions: Plasticity mechanisms and strength scaling laws

Divya Singh [a,1], Daniel S. Gianola [b], Timothy J. Rupert [a,c,*]
[a] Department of Materials Science and Engineering, University of California, Irvine, CA 92697, USA
[b] Materials Department, University of California, Santa Barbara, CA 93106, USA
[c] Department of Mechanical and Aerospace Engineering, University of California, Irvine, CA 92697, USA
[1] Present address: Engineering Department, Utah Tech University, St. George, UT 84780, USA
* Corresponding Author: trupert@uci.edu

**Abstract**

Linear complexions are stable defect states, where the stress field associated with a dislocation induces a local phase transformation that remains restricted to nanoscale dimensions. As these complexions are born at the defects which control plasticity in metals, it is crucial to understand their impact on subsequent mechanical properties. In this work, atomistic modeling is used to understand how dislocation mechanics are altered by the presence of nanoparticle array linear complexions in a Ni-Al alloy. Molecular dynamics simulations are used to identify the critical shear stress needed to drive dislocation breakaway, first for nanoparticle arrays formed by Monte Carlo/molecular dynamics methods to represent realistic configurations and subsequently for simplified models that allow the effects of particle spacing and size to be varied in a controlled manner. A combined bowing and progressive unpinning mechanism is uncovered, leading to the demonstration of a new strength scaling law that differs in keys ways from classical Orowan bowing.

**Keywords:** Plasticity, dislocations, nickel alloys, molecular dynamics (MD), linear complexions

# 1. Introduction

Defects in crystalline materials have unique local structure, chemistry, and stress state that differ from the bulk, which can drive local complexion transitions. While complexions have been explored for grain boundaries and interfaces for many years (see, e.g., Refs. [1, 2]), recent studies have shown that dislocations can also host novel local states. Kuzmina et al. [3] were the first to observe such linear complexions, finding evidence of face-centered cubic austenite regions localized near edge dislocations in an otherwise body-centered cubic Fe-Mn alloy. These authors found that the local chemical compositions near the dislocation agreed with the equilibrium Mn concentrations from the bulk phase diagram associated with this alloy, meaning that stress-driven segregation brought this local region into the right conditions for a localized phase transition to occur. Kwiatkowski da Silva et al. [4] subsequently showed that these Mn-enriched regions near the dislocation did not grow under extended isothermal tempering, emphasizing their stability within the microstructure. Similar linear complexions comprised of nanoparticle arrays were observed in low alloy reactor pressure vessel steels by Odette et al. [5].

Linear complexions adhere to rigorous thermodynamic theory, provided that the dimensionality of confinement is accounted for. For example, Frolov and Mishin [6] derived a set of thermodynamic descriptions for one-, two-, and three-dimensional phases that could be used to build expressions such as adsorption and phase coexistence equations. More recently, Korte-Kerzel et al. [7] proposed that this thermodynamically-stable complexion or "defect phase" concept could be used to build defect phase diagrams that would enable the design of advanced materials. While linear complexions were first observed in body-centered cubic alloys, recent research has demonstrated analogous defect states in face-centered cubic alloys as well. Zhou et al. [8] observed solute decoration patterns near collections of dislocations in a Pt-Au alloy, with much higher dopant segregation levels than the Cottrell cloud theory would suggest. Turlo and Rupert [9] used atomistic modeling to probe the variety of linear complexion types that would be expected in face-centered cubic alloys, categorizing the complexions according to their effect on the dislocation core. These authors showed that the highly segregated states exhibited pronounced chemical ordering, again differentiating linear complexions from Cottrell [10] and Suzuki [11] segregation.

While interesting by themselves as confined nanoscale phases, linear complexions should also act as strong obstacles to dislocation motion and provide a strengthening effect. A static strain aging effect was indeed observed by Kwiatkowski da Silva et al. [4] in the Fe-Mn system, demonstrating that the dislocations needed to break away from internal obstacles to initiate plastic flow. While Cottrell atmospheres of various dopants can also result in yield strength increases and strain aging [12, 13], the segregation is in the form of a solid solution region rather than the chemically ordered patterns found at linear complexions. The yield strength of engineering alloys is fundamentally determined by the difficulty of moving dislocations past obstacles within the microstructure. For example, precipitates can restrict dislocation motion by disrupting the migration path for the defect, with either the cutting of coherent particles [14] or bowing around incoherent particles [15] required for plasticity to commence. Linear complexions are similar to bulk precipitates in that both are particles with distinct structural and/or chemical states, yet linear complexions entail the localization of the obstacles along the dislocation line a priori, rather than random distribution within the bulk. Evidence of strengthening was obtained using molecular dynamics (MD) simulations by Turlo and Rupert [16] for a body-centered cubic Fe-Ni alloy, with linear complexions found to have much higher strength than solid solution configurations with similar dopant concentrations. A similar linear complexion strengthening effect in Fe-based alloys containing Cu, Ni, Mn, Si and P dopants was observed by Pascuet et al. [17]. A general strengthening effect is expected for face-centered cubic alloys as well, which Singh et al. [18] demonstrated with MD simulations for a collection of face-centered cubic alloys. However, the atomic-scale details of plasticity in the presence of linear complexions have not been studied for face-centered cubic alloys. In particular, the critical deformation mechanisms associated with dislocation motion and the expected scaling of alloy strength with microstructural variation must be uncovered.

In this study, the strengthening effect of nanoparticle array linear complexions is investigated in the Ni-Al system using atomistic simulations. The effect of increasing alloy concentration is first investigated, showing that larger complexion particles and more complete decoration of the dislocation line leads to an enhanced strengthening effect. The critical deformation mechanisms associated with dislocation breakaway from nanoparticle arrays are isolated, showing that dislocation bowing occurs yet progressive unpinning of the dislocation from the highly stressed region near the particles is also required for motion. Individual nanoparticle complexions models are then used to investigate the effect of particle size and spacing in a systematic manner, leading to the formulation of a modified strength scaling law. The unpinning mechanism means that the full distance between particles, not just the open gap between neighboring particles, and the particle size must be considered to accurately model strengthening from nanoparticle array linear complexions.

## 2. Computational Methods

All atomistic simulations were carried out using the Large-scale Atomic/Molecular Massively Parallel Simulator (LAMMPS) [19], while atomic scale analysis and visualization was performed with OVITO [20] using the Polyhedral Template Matching (PTM) [21] and Dislocation Extraction Algorithm (DXA) [22] methods. PTM was used to analyze the local structural ordering and enable phase identification while DXA was used to analyze dislocation structure. When necessary, manipulation of atomic datasets was accomplished with the open-source code ATOMSK [23]. An embedded atom method (EAM) interatomic potential for Ni-Al [24] was used that is capable of reproducing important features of this alloy system, such as the major stable phases on the experimental phase diagram and important defect energies. It is essential that the potential be able to accurately represent the structure and chemistry of equilibrium phases in this system. First, local equilibrium configurations were created with a hybrid Monte Carlo (MC)/MD method, following the procedures described in detail by Turlo and Rupert [9]. In short, structural relaxations were accomplished with MD, while a variance-constrained semi-grand canonical ensemble was used for the MC portion to target a specific composition following the method of Sadigh et al. [25]. Each starting simulation cell had periodic boundary conditions in all directions and contained two pairs of Shockley partial dislocations, relaxed from two starting edge dislocations, with the alloying element concentration of the Ni-Al alloys varied to give a range of nanoparticle decoration states at the dislocations. The simulations were performed at 300 K and zero pressure under an NPT ensemble, with MC/MD terminated when the corresponding energy gradient for the last 20 ps is less than 0.1 eV/ps.

The procedure described above is able to find the local structure of linear complexions, yet the two pairs of dislocation pairs can interact with one another through their stress fields due to the limited size of MD simulation cells. To enable a focused study of dislocation mechanics, half of the simulation cell was removed, leaving only one pair of partial dislocations and its associated nanoparticle array, with an example shown in Fig. 1(a). Fig. 1(b) presents a view from the bottom up, showing that Al segregates along the dislocation lines and causes the formation of confined Ni$_3$Al nanoparticles with an L1$_2$ ordered structure. Non-periodic shrink wrapped boundary conditions were used in the direction perpendicular to the glide plane while periodic boundary conditions were applied to the other two directions. A constant shear stress was applied by adding appropriate per atom forces on 1 nm slabs at the top and bottom of the simulation cell, while the canonical ensemble (NVT) was applied to all other atoms. The simulation cell was held at the target shear stress value for 100 ps. If the dislocation did not move away from the starting position, the shear stress was determined to be sub-critical and this value was increased until motion occurred and the critical shear stress for dislocation breakaway could be identified. This simulation methodology is widely used in the literature to identify critical shear stresses for dislocation motion and to study dislocation mobility as a function of applied stress. For example, Yin et al. [26] recently reported the effect of chemical short-range order in refractory high-entropy alloys on dislocation mobility using this method.

A second set of simulations was performed on isolated nanoparticles chosen from the linear complexion, to systematically vary the nanoparticle size and spacing. The specific values used for these variables and the rationale for these choices will be presented in the Results and Discussion section, as their selection was motivated by early simulation results. Therefore, two sets of virtual experiments were run, with deformation applied to (1) models containing an array of nanoparticles with statistical variations in particle size and spacing representative of realistic linear complexion configurations and (2) idealized models containing a single particle so that size and spacing could be varied in a controlled manner.

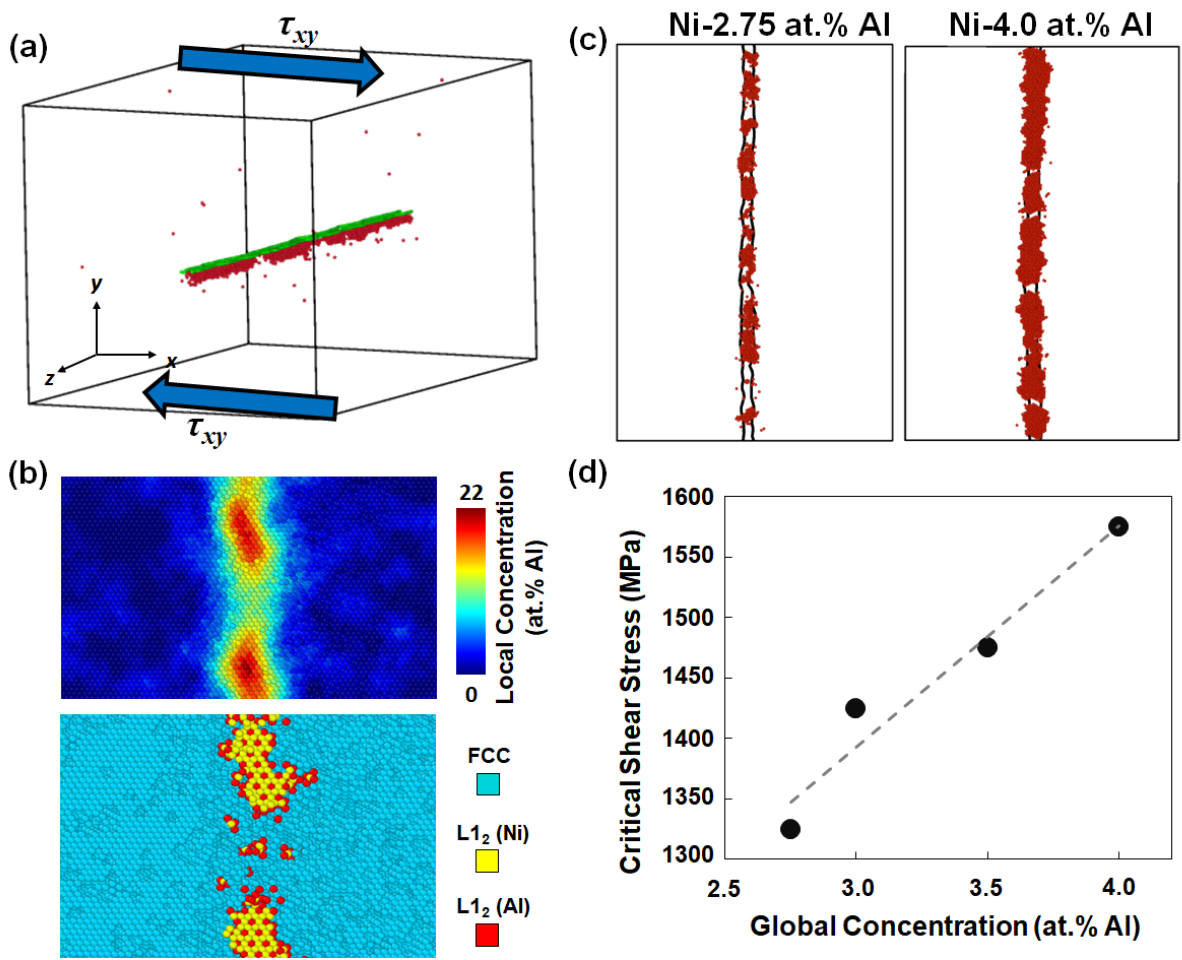

Figure 1. (a) The initial simulation cell used in this work, consisting of a pair of Shockley partial dislocations (green) and a nanoparticle array linear complexion (red to denote $L1_2$ structure as determined by PTM) under a constant applied shear stress. (b) Atomic snapshots of local Al concentration and structural type near two complexion particles from within the array, showing segregation and a local transformation to $L1_2$ structure. (c) Bottom-up views (+Y-direction) of nanoparticle array linear complexions from two representative Ni-Al samples, where increasing Al concentration causes growth of the complexion particles and increased coverage of the dislocation line. (d) Critical shear stress values as a function of global Al concentration, with a dashed grey line provided to demonstrate the general trend and guide the eye.

## 3. Results and Discussion

Fig. 1(c) shows the samples with the lowest and highest concentration of Al, where increased alloying leads to larger complexions and a reduction in the spacing between them. The growth of the complexions is most notable in the length direction, along the dislocation lines, but a slight thickening effect normal and perpendicular to the dislocation lines is also observed. The critical shear stress was extracted from a number of samples between these conditions, to understand the strengthening effect of nanoparticle array linear complexions. These critical shear stress values are presented in Fig. 1(d), where they are plotted as a function of the Al concentration in the system. A rapid increase in the stress needed to move the dislocations is observed, reaching values of ~1575 MPa for the highest Al concentration where large nanoparticles almost completely decorate the dislocations. Such a strength is very high for Ni-rich alloys, exceeding that of many advanced engineering alloys. For example, Inconel alloy 718 has a room temperature yield strength of 1167 MPa in the precipitation strengthened condition [27], which translates into a 675 MPa maximum shear strength using a von Mises yield criterion. A dashed line is provided in Fig. 1(d) to guide the eye and this linear fit works reasonably well to reproduce the trend of the data, which will be discussed later. Worth noting is that a linear dependence of strength on alloying element concentration is much stronger than the typical square root dependence of strength on composition found for solid solution effects.

The key dislocation breakaway mechanisms observed in these simulations are presented in Fig. 2 for the Ni-3.5 at.% Al specimen, with times, $t$, denoted below each image. We generally observed that the pair of partial dislocations moved together, meaning their collective motion can be considered together rather than requiring separate analysis for the leading and trailing partial. The dislocation first starts to move within a gap between particles P4 and P5 (marked with green lettering) in Fig. 2(a). This gap is large relative to the others observed in this array, with the locations of the local dislocation-particle restrictions at $t = 7.5$ ps marked with orange arrows in the inset. At first glance, this bowing out appears similar to the classical Orowan bowing observed in the presence of bulk precipitates, where the applied shear stress works to overcome the line tension of the dislocations to create a curved configuration. However, there are fundamental differences between the two that must be acknowledged. First, the linear complexion particles lie below the dislocation slip plane, unlike in the case of traditional precipitate obstacles where they lie in the slip plane. The linear complexion particles do not provide a physical barrier that blocks the pathway of the dislocations, but rather restrict motion by interacting with the stress fields of the dislocations. The choice of the largest gap in the particle array is consistent with Orowan bowing, as larger distances between restricting obstacles make the dislocations easier to bow out owing to smaller line tension.

Once bowing occurs, the dislocations progressively unzip from the pinning particles, as shown in Fig. 2(b). In these images, specific sites where the dislocations pull away from the particle in the time elapsed (1 ps) between the two images are marked with blue arrows. The dislocations unzip quickly from particle P4 between Figs. 1(a) and (b), spending more time stuck in a restricted configuration as they unzip from particles P3 and P5. Both of the latter particles are much larger than the former, showing that particle size is important for its ability to resist the unzipping mechanism. Thus, the two predominant factors affecting strengthening in the presence of nanoparticle arrays linear complexion appear to be (1) particle spacing and (2) particle size. It is important to note that both bowing and unzipping must occur sequentially for the dislocations to move, meaning the original gap between particles is not enough by itself to determine the strength of the obstacle. In fact, stable and sessile bowed configurations at empty gaps were observed for sub-critical applied stress values. The distance along the particles where the dislocations are pinned is important as well, meaning the dislocations must bow out further into a wider arc to accommodate this feature.

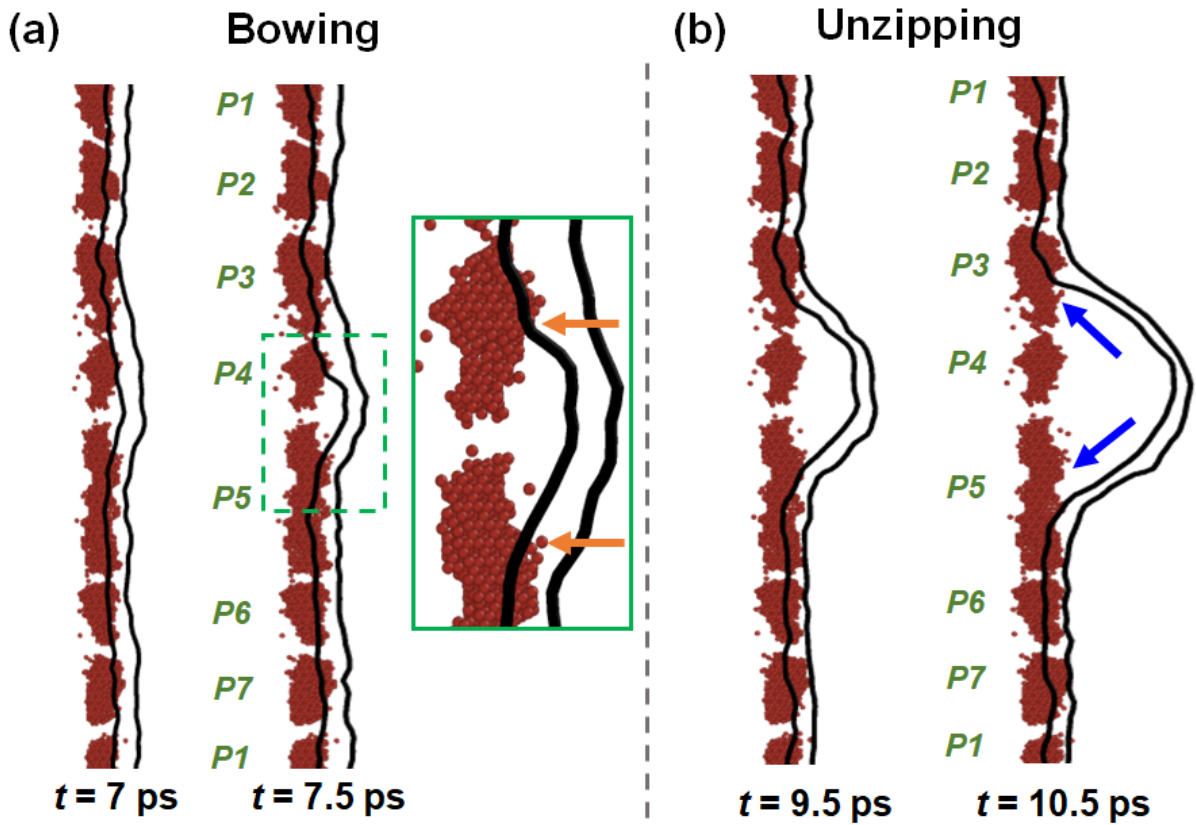

**Figure 2.** Mechanisms associated with dislocation breakaway from nanoparticle array linear complexions in Ni-Al. (a) The partial dislocation pair first bows outward in regions of open space, where defects are not restricted by a particle and its associated stress field. The inset provides a closer view of the initial bowing location for this configuration, with dislocation-particle pinning points denoted by orange arrows. (b) The dislocations pull away from the locations where they were still pinned at the nanoscale particles at $t$ = 9.5 ps (denoted by blue arrows) through a progressive unzipping motion. The dislocations unzip very quickly from smaller particles, such as particle P4.

While the equilibrium linear complexion configurations shown in Figs. 1 and 2 demonstrate a strong strengthening effect and uncover key mechanisms, the presence of a variety of particle sizes and different individual gap sizes between them makes it challenging to develop a systematic understanding of these variables. A second set of model simulation cells containing one particle each and different lengths in the dislocation line length direction were used to systematically vary particle size and spacing based on the mechanistic insights obtained above. Small, medium, and large particles with sizes ($h$) of 3.2 nm, 7.4 nm, and 11 nm were chosen to sample the variety of particles observed in Figs. 1 and 2. We note that the particle size is described in terms of its length along the dislocation line, as this is the length scale over which the unzipping mechanism occurs. These three particles were chosen from the MC/MD configurations, with the other particles removed to make simulation cells with different lengths in the Z-direction. With only one particle per simulation cell and periodic boundary conditions, the box length therefore sets the spacing between the center of mass of each particle. Particle spacing was varied from approximately 4 to 45 nm. Two sets of examples are depicted in Fig. 3(a) for reference, depicting the small and medium particle sizes, with a small particle spacing in the upper row (particles are nearly touching) and a large spacing used in the bottom row.

The critical values of applied shear stress needed to induce dislocation breakaway were again identified. The left side of Fig. 3(b) shows the critical dislocation configurations associated with (1) the starting state, (2) the bowing mechanism, and (3) the unzipping mechanism, demonstrating that the key details of breakaway are the same in these model system as they were in the more realistic linear complexion configurations. The right side of Fig. 3(b) shows the distribution of atomic hydrostatic stresses on a plane through the linear complexion particles. These images show that the hydrostatic stress fields of the dislocations (tensile) and the particle (compressive) largely cancel one another out in the starting configuration, with the resulting compressive stress from the particle more clearly observed once the dislocations start to move away in the middle and bottom frames. This compressive stress field is responsible for the pinning that necessitates unzipping from the nanoparticle.

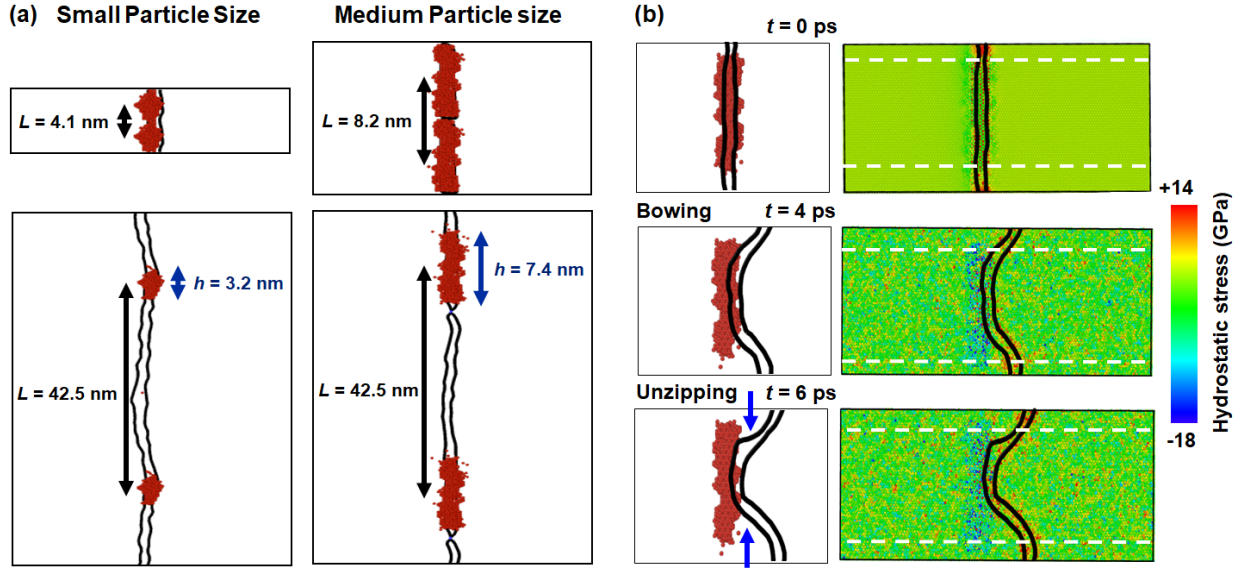

**Figure 3.** Simplified simulation cells allow important variables such as linear complexion particle size and spacing to be changed systematically. (a) Example configurations with small and large spacings, for both small and medium sized particles. (b) Atomistic configurations and hydrostatic stresses associated with the starting configuration, as well as the bowing and unzipping mechanisms associated with dislocation breakaway (direction of unzipping is denoted by blue arrows). A region of compressive stress associated with the complexion particle (whose bounds are marked by white dashed lines) is observed in the middle of the images and resists dislocation motion, requiring unzipping.

The compiled data for the critical shear stresses for complexions of different sizes is plotted in Fig. 4(a) as a function of particle spacing. For a given particle size, reduction in the particle spacing leads to a rapid increase in the strength of the obstacle. A smaller particle spacing means that the dislocation bowing is much more difficult, as the dislocation line must deform into a semi-circle configuration with a tighter radius. The effect of particle size is also clearly observed in this data, with larger particles pushing the data up and to the right, meaning that a higher obstacle strength is achieved for the same particle spacing.

Most interesting is to understand how the data from the nanoparticle array linear complexion strengthening deviates from prior observations for traditional precipitation strengthening. When particles block the motion of dislocations and bowing is necessary, strength typically scales according to the classical Orowan equation [28]:

$$\tau_{crit.} = \alpha \cdot \frac{Gb}{(L-2r)} \tag{1}$$

where $\tau_{crit.}$ is the critical shear stress for a dislocation to overcome the precipitate obstacle, $\alpha$ is a pre-factor associated with the line tension of the dislocation, $G$ is the shear modulus of the crystal through which the dislocation moves and bows, $b$ is the Burgers vector of the lattice, $L$ is the particle spacing (taken as the distance from particle center to particle center), and $r$ is the particle radius. The pre-factor $\alpha$ is related to the dislocation character, so would be constant provided the same type of dislocation is studied, as was done here. Equation 1 indicates that the main consideration of the Orowan model is the open space between the particles, obtained by subtracting $2r$ from $L$, making this model primarily geometric. The segment of the dislocation line which hits the particle remains as an Orowan loop after the dislocations bow around the particles and move away, so it does not participate in the bowing mechanism. Particle size only comes into play as it affects the open gap, and the particles do not directly provide any strengthening increment by restricting dislocation motion. The top image of Fig. 4(b) schematically shows the important features of the Orowan model.

The important features of nanoparticle array linear complexion strengthening are shown in the bottom frame of Fig. 4(b), emphasizing a number of unique points. First, while the bowing starts at open gaps, it continues during the unzipping process. This means that the relevant length scale for bowing is not the open gap length, but rather the entire particle spacing, measured from the center-of-mass of one particle to another. In addition, the particle size is important because larger particles make the unzipping process more difficult. This was shown anecdotally in Fig. 2 but is also quantitatively corroborated in the data from Fig. 4(a). Put together, one can build a modified strength scaling law for nanoparticle array linear complexions:

$$\tau_{crit.} = \beta_{size} \cdot \frac{Gb}{L} \tag{2}$$

where $\beta_{size}$ is a pre-factor associated with the particle size and its effect on strength and the other variables hold the same meaning as earlier presented. This new strength scaling law was used to fit the measured critical shear stress values from this study, using values of *G* and *b* for pure Ni as these describe the matrix away from the linear complexion, with Fig. 4(a) showing that Equation 2 can accurately describe the strength of nanoparticle array linear complexions. While not shown here, Equation 1 resulted in very poor fits to the data from this study, as the (*L* – 2*r*) scaling does not work well and there was no way to account for different obstacle strength associated with particle size. Interestingly, the rapid increase in strength as obstacle spacing is decreased to very small values is also more meaningful than a similar trend in Orowan bowing. For traditional precipitation strengthening, the relevant spacing going to zero would mean that the entire material is now comprised of the precipitate phase, making the question of how difficult it is to move a dislocation through the Al matrix nonsensical. In contrast, *L* decreasing until it equals *h* simply implies that the nanoparticle linear complexions completely cover the dislocation lines, which is possible and in fact nearly occurs in Fig. 1(c). The curve fits in Fig. 4(a) are therefore cut off at *L* = *h*, with all three curves terminating at critical shear stress values above 2000 MPa. Hence, extreme strengthening to the limits of the theoretical interpretation is actually accessible with nanoparticle array linear complexions. The pre-factor $\beta_{size}$ that best fits each of the three data sets for different particle sizes becomes larger as particle size increases, with the values plotted in Fig. 4(c) as a function of particle size. While a physical meaning of the fitting form is not clear at this point, a linear fit works well for describing the $\beta_{size}$ values for our data, which represents a range of sizes from very small to very large.

    Recalling that the critical shear stress data shown in Fig. 1(d) also followed a linear trend, the strengthening observed in the realistic nanoparticle array linear complexion models is

dominated by the fact that the particles get larger as Al concentration increases. For both the realistic and simplified simulation models studied in this work, high strengths in the GPa-range are achievable with different combinations of particle size and spacing. This is likely a consequence of an innate feature of linear complexions: obstacles are created directly at the dislocation. This means that for a given volume fraction of particles in a material, those in a linear complexion are efficiently placed, allowing for the particle spacing to be much smaller than if the particles were randomly placed throughout the matrix. For example, for the linear complexion configuration shown in Fig. 2, the particle spacing values are in the range of 4-8 nm, much smaller than the tens to hundreds of nanometer spacings typically observed in traditional precipitation strengthened Al alloys (see, e.g., Refs. [29-31]). It is also useful to consider this small particle spacing in the context of the data and strength scaling laws shown in Fig. 4. Nanoparticle spacings of 4-8 nm entered into Equation 2 would give strengths in the GPa-range, consistent with the magnitude of the strength values found in Fig. 1(d). However, it is not clear what exact value (e.g., the mean or some other descriptor of the particle size and spacing distribution) should be entered into Equation 2 to describe the more complex collection of particles in the realistic linear complexion samples, marking an area for future inquiry. Finally, we note that nanoparticle array linear complexions contain many particles in close proximity, meaning that complex bowing mechanisms may be active in some cases. For example, Xu et al. [32] showed that collinear obstacles along a long particle array can influence dislocation bowing mechanisms, although these authors found that intermediate obstacles did not influence the critical shear stresses.

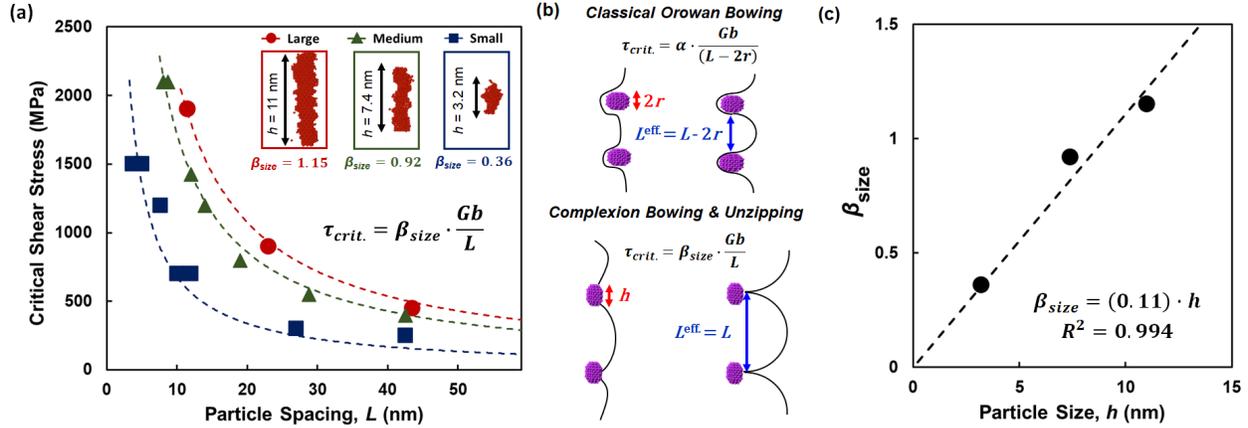

**Figure 4.** (a) Compiled data for the critical shear stress needed to induce dislocation breakaway in Ni-Al for linear complexions of different sizes (Small = 3.2 nm long, Medium = 7.4 nm long, Large = 11 nm long) and particle spacings, taken as the distance between the centers of mass. (b) Schematic representations of classical Orowan bowing, as compared to linear complexion bowing and unzipping, with important material length scales noted for each. (c) Values of the pre-factor for the nanoparticle linear complexion strengthening model plotted as a function of particle size, showing a linear relationship.

## 4. Conclusions

In this work, atomistic modeling is used to uncover the mechanics of dislocation breakaway from nanoparticle array linear complexions in a Ni-Al alloy, with key mechanisms of sequentially bowing and unzipping identified and a strength scaling law created. While Orowan bowing occurs because the path of the dislocations is physically blocked by a precipitate, linear complexions interact with the dislocations through their hydrostatic stress fields. Bowing starts to pull the dislocations away from the linear complexion particles, but an unzipping process is also required to achieve full breakaway. A systematic study of dislocation motion near these unique obstacles establishes the factors affecting strengthening, namely the size of the particles and the entire spacing, defined by the center-to-center distance. With these mechanisms in mind, a new strength scaling law is proposed that can accurately describe the strengthening effect of nanoparticle array linear complexions of a wide range of particle sizes and particle spacings. As a whole, extreme strengthening effects are observed for nanoparticle array linear complexions, marking these features as promising targets for next-generation microstructural engineering.

**Acknowledgements**

This research was sponsored by the U.S. Army Research Office under Grant Number W911NF-21-1-0288. The views and conclusions contained in this document are those of the authors and should not be interpreted as representing the official policies, either expressed or implied, of the Army Research Office or the U.S. Government. The U.S. Government is authorized to reproduce and distribute reprints for Government purposes notwithstanding any copyright notation herein.

**Declaration of interest**

The authors declare that they have no known competing financial interests or personal relationships that could have appeared to influence the work reported in this paper.